\documentclass[wssquare]{ws-procs961x669}

\pdfoutput=1
\usepackage{mathtools,mathrsfs}
\usepackage{tensor}

\makeatletter
\newcommand*{\defeq}{\mathrel{\rlap{%
	\raisebox{0.3ex}{$\m@th\cdot$}}%
	\raisebox{-0.3ex}{$\m@th\cdot$}}%
	=}
\newcommand*{\eqdef}{=\mathrel{\rlap{%
	\raisebox{0.3ex}{$\m@th\cdot$}}%
	\raisebox{-0.3ex}{$\m@th\cdot$}}%
	}
\makeatother

\def\pad{{\partial}}

\def\sg{\textsl{g}}

\def\rP{\mathrm{P}}
\def\rS{\mathrm{S}}
\def\vS{v_\mathrm{S}}

\def\cO{\mathcal{O}}

\def\eK{\mathcal{K}}
\def\eR{\mathcal{R}}

\def\maT{\mathfrak{T}}

\usepackage[caption=false]{subfig}
\newcommand{\subfigimg}[3][,]{%
  \setbox1=\hbox{\includegraphics[#1]{#3}}
  \leavevmode\rlap{\usebox1}%
  \rlap{\hspace*{0pt}\raisebox{.5\baselineskip}{\small{#2}}}%
  \phantom{\usebox1}
}

\usepackage{xcolor}

\usepackage{hyperref}
\hypersetup{colorlinks,linkcolor={blue!55!black},citecolor={red!45!black},urlcolor={blue!45!black},breaklinks=true}

\begin{document}

\title{Physical black holes in semiclassical gravity}

\author{Sebastian Murk}
\address{Department of Physics and Astronomy, Macquarie University,\\
Sydney, New South Wales 2109, Australia\\
and\\
Sydney Quantum Academy,\\
Sydney, New South Wales 2006, Australia\\
E-mail: \href{mailto:sebastian.murk@mq.edu.au}{sebastian.murk@mq.edu.au}}

\author{Daniel R.\ Terno}
\address{Department of Physics and Astronomy, Macquarie University,\\
Sydney, New South Wales 2109, Australia\\
E-mail: \href{mailto:daniel.terno@mq.edu.au}{daniel.terno@mq.edu.au}}

\begin{abstract}
We derive and critically examine the consequences that follow from the formation of a regular black or white hole horizon in finite time of a distant observer. In spherical symmetry, only two distinct classes of solutions to the semiclassical Einstein equations are self-consistent. Both are required to describe the formation of physical black holes and violate the null energy condition in the vicinity of the outer apparent horizon. The near-horizon geometry differs considerably from that of classical solutions. If semiclassical physics is valid, accretion into a black hole is no longer possible after the horizon has formed. In addition, the two principal generalizations of surface gravity to dynamical spacetimes are irreconcilable, and neither can describe the emission of nearly-thermal radiation. Comparison of the required energy and timescales with established semiclassical results suggests that if the observed astrophysical black holes indeed have horizons, their formation is associated with new physics.
\end{abstract}

\keywords{black holes; white holes; general relativity; semiclassical gravity; quantum aspects of black holes; energy conditions; surface gravity.\\[3mm]
Contribution to the proceedings of the 16th Marcel Grossmann meeting (5--10 July 2021).\\
Session classification: Theoretical and observational studies of astrophysical black holes}

\bodymatter
\thispagestyle{empty}

\section{Introduction} \label{sec:intro}
Our current understanding of ultra-compact objects (UCOs) can be summarized as follows: the existence of astrophysical black holes (ABHs) --- dark massive compact objects --- is established beyond any reasonable doubt. However, it is unclear when, how, or if at all these UCOs develop the standard black hole attributes, such as horizons and singularities. A large number of models, often with deliberately designed features (or lack thereof), purport to describe ABHs. Translating the differences between these models into potentially observable differences between the signals received from the black hole candidates they describe is one of the most exciting topics in gravitational physics research \cite{map:19,cp:19}.

There is no unanimously agreed upon definition of a black hole \cite{c:19}. The salient feature of a mathematical black hole (MBH) \cite{f:14} is the event horizon that separates our outside world from an inaccessible interior. However, event horizons are global teleological entities that are --- even in principle --- unobservable \cite{v:14,cp:17}, and observational, numerical, and theoretical studies focus on other characteristics of black holes. A trapped region is a spacetime domain where both ingoing and outgoing future-directed null geodesics emanating from a spacelike two-dimensional surface with spherical topology have negative expansion \cite{he:book:73,fn:book:98,f:book:15}. Its evolving outer boundary is the apparent horizon. This definition captures the most fundamental feature of black holes as spacetime regions that nothing, not even light, can escape. It is also local, and thus physically observable: the escape is not possible now, but the notion of ``now'' depends on the observer. Following the characterization scheme of Ref.~\citenum{f:14}, we refer to a trapped spacetime region {that is bounded by an apparent horizon} as a physical black hole (PBH) \cite{mmt:rev:21}.

Although it cannot be observed, the event horizon is a useful asymptotic concept, and in many situations it is reasonable to assume that MBHs provide a good description of certain aspects of ABHs. Indeed, the predictions of various alternatives to black holes are most often compared with the standard Schwarzschild/Kerr paradigm \cite{map:19,cp:19}. However, using this asymptotic concept at finite times is logically unsatisfactory, and in the situations we consider below cannot be sustained.

\begin{figure*}[!htbp]
  	\centering
 	\begin{tabular}{@{\hspace*{0.05\linewidth}}p{0.45\linewidth}@{\hspace*{0.025\linewidth}}p{0.45\linewidth}@{}}
  	\centering
		\subfigimg[scale=0.555]{(a)}{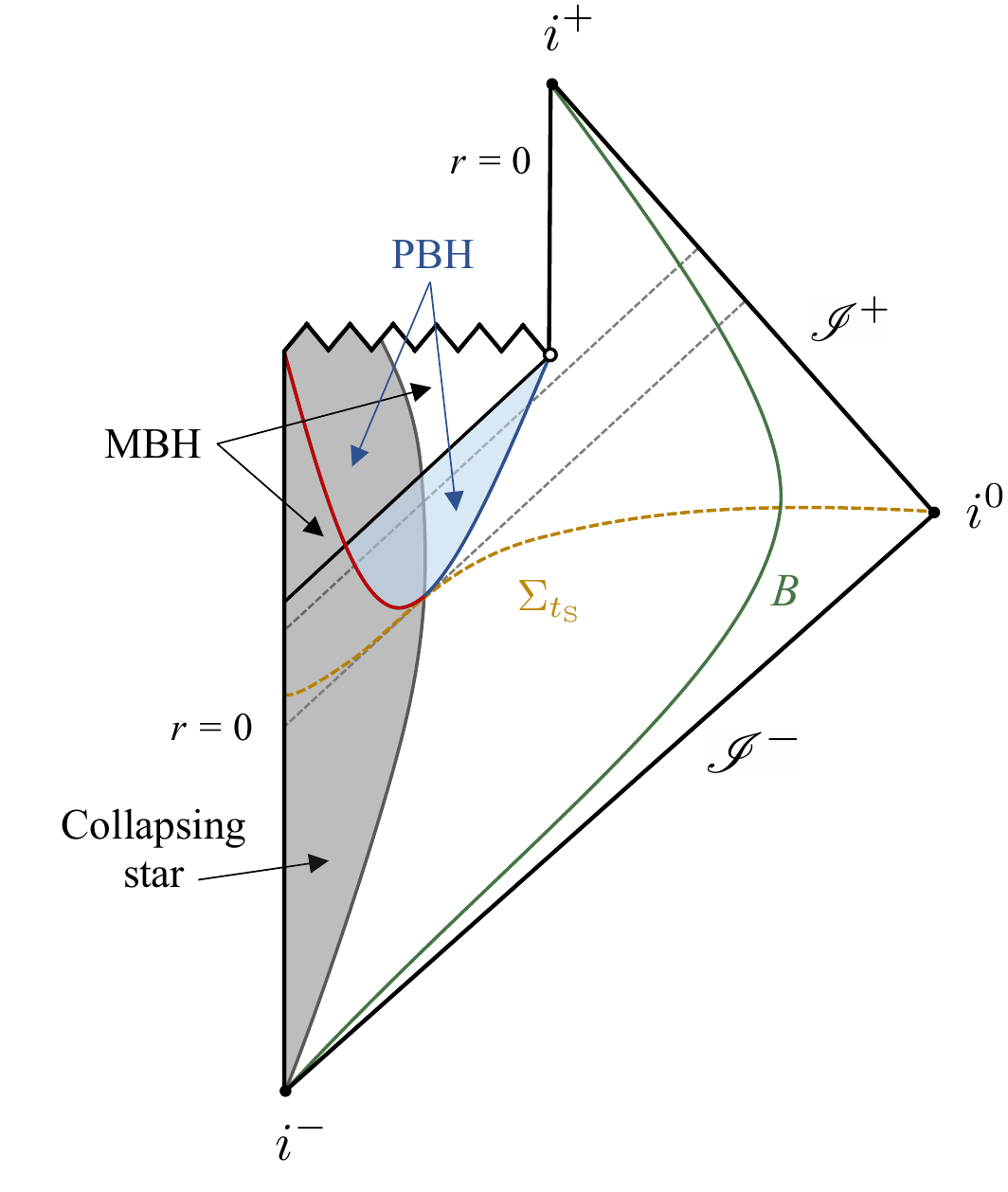} &
 		\subfigimg[scale=0.525]{(b)}{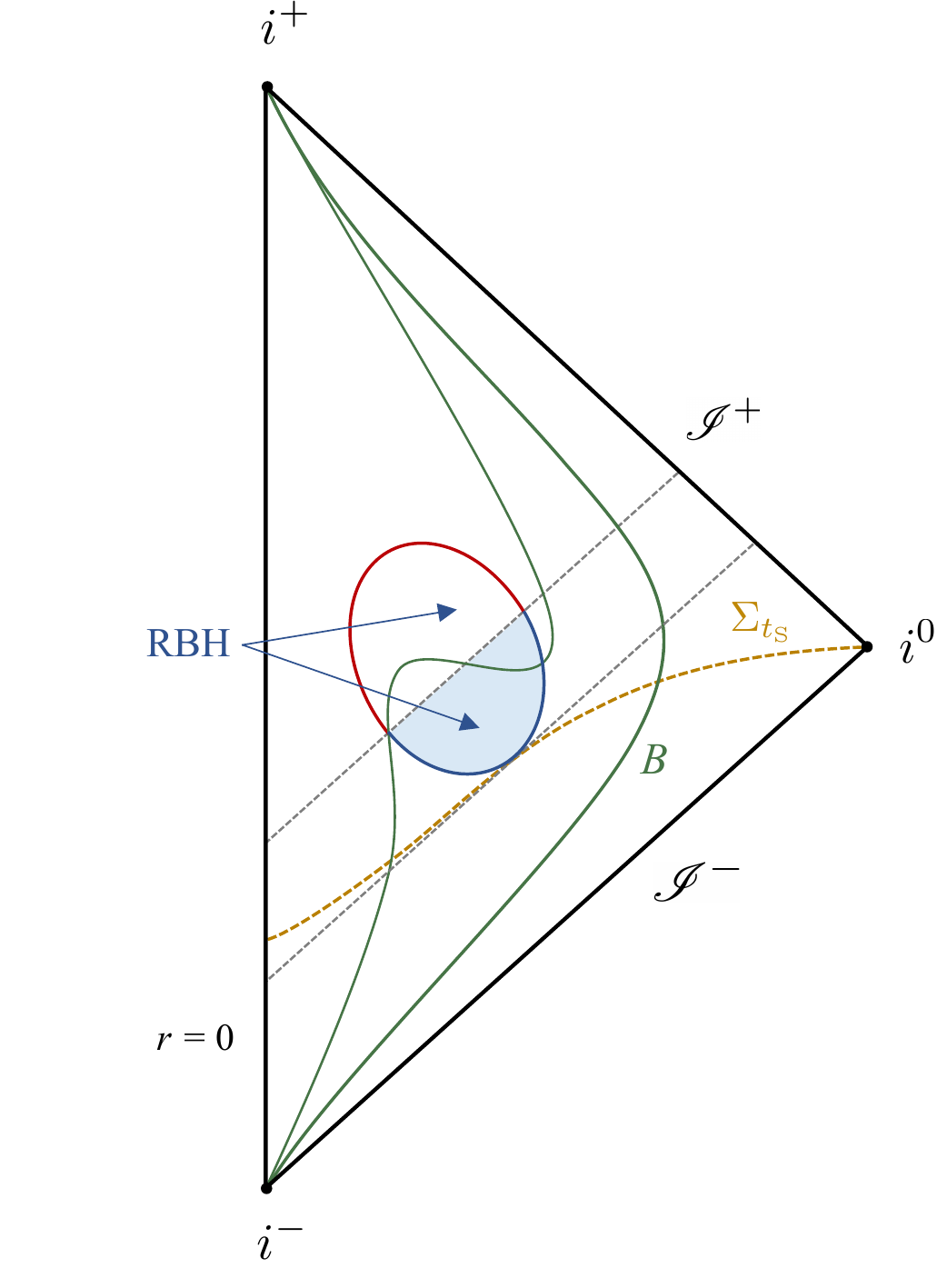}
  	\end{tabular}
	\caption{Schematic Carter\textendash{}Penrose diagram  {of the conventional formation and evaporation of (a) a black hole, and (b) a RBH. The outer apparent horizon $r_\sg(t)$ and the inner apparent horizon $r_\mathrm{in}(t)$ form the boundary of a PBH and are shown in blue and red, respectively.} The equal time surface $\Sigma_{t_\mathrm{S}}$  is drawn as a dashed orange line.  It is null at $\big(t_\mathrm{S},r_\sg(t_\mathrm{S})\big)$ and spacelike everywhere else \cite{dmt:21}. The trajectory of a distant observer Bob is indicated in green and labeled ``B''. Dashed grey lines correspond to outgoing radial null geodesics. (a) Diagrams of this type are elaborations of the original sketch by Hawking, see Fig.~5 of Ref.~\citenum{h:75}. Spacetime regions corresponding to PBH (MBH) solutions are indicated by blue (black) arrows. Light signals emitted from the quantum ergosphere \cite{y:83} (indicated by the light blue shading), i.e.\ from within the trapped region, but outside of the event horizon, reach $\mathscr{I}^+$ and are detected by Bob in his finite proper time. The collapsing matter and its surface are shown as in conventional depictions of the collapse. However, the matter in the vicinity of the outer apparent horizon $\big(t,r_\sg(t)\big)$ violates the NEC for $t\geqslant t_\mathrm{S}$. Moreover, the energy density, pressure, and flux as seen by an infalling observer Alice vary continuously across it, and the equation of state dramatically differs from that of normal matter that may have been used to model the initial EMT of the collapse. (b) The asymptotic structure of a simple RBH spacetime \cite{cdlv:20} coincides with that of Minkowski spacetime. Conditions for the smooth joining of the inner and outer apparent horizon are described in Ref.~\citenum{bhl:18}. An unmarked green line represents a hypersurface $r=\mathrm{const}$ that passes through the RBH.}
  	\label{fig:time-g}
\end{figure*}

Another question is whether the asymptotic picture is adequate to describe the relevant geometric and physical properties of black holes. We first formalize the minimal assumptions that underpin the widely used black hole notions presented in \fref{fig:time-g}, namely regularity of the apparent horizon and its finite-time formation according to the clock of a distant observer. In spherical symmetry, these two assumptions uniquely determine the formation scenario for PBHs and suffice to provide an exhaustive description of their near-horizon geometry. Its properties are the subject of this contribution. We now outline the motivation and justification for these requirements \cite{mmt:rev:21}.

In classical general relativity (GR) non-spacelike singularities destroy predictability. According to the weak cosmic censorship conjecture \cite{sg:15,c-b:09}, spacetime singularities are concealed by event horizons. Quantum gravitational effects are expected to become important when the spacetime curvature is sufficiently strong \cite{bd:book:82,fn:book:98}, i.e.\ when the Kretschmann scalar $\eK \defeq R_{\mu\nu\rho\sigma} R^{\mu\nu\rho\sigma}$ reaches the Planck scale, that is $\eK \gtrsim l_\rP^{-4}$. The near-horizon geometry is believed to be well-described by semiclassical physics. Nontrivial quantum effects are obtained within the framework of quantum field theory in curved spacetime \cite{bd:book:82}. One of its most spectacular predictions is Hawking radiation: it not only completed black hole thermodynamics \cite{fn:book:98,w:01}, but has given rise to the infamous information loss paradox \cite{h:75}. Regular (singularity-free) black holes were introduced to altogether eliminate singularities in classical gravitational collapse or as a way to resolve the problem of information loss \cite{rb:83,h:06,f:14}. Leaving aside  the interior structure of black holes, we formulate the regularity criterion as the absence of curvature singularities at the apparent horizon. A precise mathematical formulation \cite{mmt:rev:21} is provided in Sec.~\ref{metrics}.

To be considered a genuine physical object rather than merely a useful mathematical tool, the horizon must form in finite time of a distant observer, and there should be some potentially (at least in principle) observable consequences of its formation. Moreover, if black holes do indeed emit Hawking radiation, and their evolution roughly resembles that of \fref{fig:time-g}(a), then the apparent horizon forms in finite time $t_\rS$ of a distant observer. We note that, as illustrated in \fref{fig:time-g}(a), the outer apparent horizon is located outside of the event horizon \cite{h:75,fn:book:98}. Hence all signals that are emitted from the so-called quantum ergosphere \cite{y:83,bhl:18} --- part of the trapped region that lies outside of the event horizon --- reach future null infinity $\mathscr{I}^+$. \ {In addition}, models of transient (even if long-lived) regular black holes (RBHs) imply the finite-time formation and disappearance of the trapped region, as depicted in \fref{fig:time-g}(b). This leads to our second requirement: the finite-time formation of an apparent horizon according to a distant observer.

Working in the framework of semiclassical gravity, we use classical notions (horizons, trajectories, etc.) and describe dynamics via the Einstein equations $\tensor{G}{_\mu_\nu} = 8 \pi \tensor{T}{_\mu_\nu}$. We do not assume any specific matter content nor a specific quantum state $\omega$ that produces the expectation values of the energy-momentum tensor (EMT) $\tensor{T}{_\mu_\nu} \defeq \langle \tensor{\hat{T}}{_\mu_\nu} \rangle_\omega$. Note that this EMT describes the total matter content --- both the original collapsing matter and the produced excitations. We do not assume the presence of Hawking-like radiation, an event horizon, or a singularity. To simplify the exposition we work in asymptotically flat spacetimes, even if it is not essential for the resulting near-horizon geometries.

\section{Spherically symmetric solutions} \label{metrics}
It is convenient to use Schwarzschild coordinates to impose the two assumptions described above. Their coordinate singularities allow us to identify the admissible solutions. A general spherically symmetric metric in four spacetime dimensions is given by
\begin{align}
	ds^2=-e^{2h(t,r)}f(t,r)dt^2+f(t,r)^{-1}dr^2+r^2d\Omega_2. \label{sgenm}
\end{align}
These coordinates provide geometrically preferred foliations with respect to Kodama time, a natural divergence-free preferred vector field \cite{f:book:15,av:10}. In asymptotically flat spacetimes  the time $t$ represents the proper time  of a distant static observer. Using the advanced null coordinate $v$, the metric is written as
\begin{align}
  ds^2=-e^{2h_+}\left(1-\frac{C_+}{r}\right)dv^2+2e^{h_+}dvdr +r^2d\Omega_2 . \label{lfv}
\end{align}
The Misner--Sharp (MS) mass $C(t,r)/2$ \cite{f:book:15,ms:64} is invariantly defined via
\begin{align}
	f(t,r) \defeq 1-C/r \defeq \partial_\mu r\partial^\mu r, \label{defMS}
\end{align}
and thus $C(t,r)\equiv C_+\big(v(t,r),r\big)$. The functions $h(t,r)$ and $h_+(v,r)$ play the role of integrating factors in coordinate transformations, such as
\begin{align}
	dt=e^{-h}(e^{h_+}dv- f^{-1}dr). \label{intf}
\end{align}
The apparent horizon is located at the Schwarzschild radius $r_\sg(t)\equiv r_+(v)$ that is the largest root of $f(t,r)=0$ \cite{f:book:15,fefhm:17}. In $(v,r)$ coordinates the tangents to ingoing and outgoing radial geodesics are given by
\begin{align}
	l_{\mathrm{in}}^\mu=(0,-e^{-h_+},0,0), \qquad l_{\mathrm{out}}^\mu=(1,\tfrac{1}{2} e^{h_+}f,0,0), \label{null-v}
\end{align}
respectively. They are normalized to satisfy $l_{\mathrm{in}}\cdot l_{\mathrm{out}}=-1$, and their corresponding expansions are
\begin{align}
	\vartheta_{{\mathrm{in}}} = - \frac{ {2} e^{-h_+}}{r} , \qquad \vartheta_{{\mathrm{out}}}=\frac{e^{h_+}f}{r} ,
\end{align}
thus identifying the domain $r \leqslant r_+\equiv r_\sg$ as a PBH.

Using the retarded null coordinate $u$, the metric is written as
\begin{align}
  ds^2 = - e^{2h_-} \left( 1 - \frac{C_-}{r} \right) du^2 - 2e^{h_-}du dr + r^2 d\Omega_2 . \label{lfu}
\end{align}
In $(u,r)$ coordinates the tangents to ingoing and outgoing radial geodesics are given by
\begin{align}
	l_{\mathrm{in}}^\mu = (1,-\tfrac{1}{2} f(u,r)e^{h_-(u,r)},0,0), \qquad l_{\mathrm{out}}^\mu = (0,e^{-h_-(u,r)},0,0),
\end{align}
respectively. Again, the tangents are normalized by $l_{\mathrm{in}}\cdot l_{\mathrm{out}}=-1$, and their respective expansions are given by
\begin{align}
	\vartheta_{{\mathrm{in}}} = - \frac{e^{h_-}f}{r}, \qquad \vartheta_{{\mathrm{out}}}=\frac{2e^{-h_-}}{r} .
	\label{eq:ur-tangent-expansions}
\end{align}
As a result, the domain $r\leqslant r_- \equiv r_\sg$ is a white hole and the Schwarzschild radius is the location of the anti-trapping horizon. As in the case of the maximally extended Schwarzschild solution, these two scenarios describe different physical settings.

The Schwarzschild coordinates become singular as $r \to r_\sg$. We extract information about the EMT and therefore about the near-horizon geometry by studying how various divergences cancel to produce finite curvature scalars. Singular points are identified through the presence of incomplete geodesics in their vicinity and are excluded from manifolds representing the spacetime. These geodesics are inextendible in at least one direction, but the range of their generalized affine parameter (proper time for timelike geodesics) is bounded \cite{p:68,he:book:73,c-b:09,tce:80}. We focus on curvature singularities and formalize the regularity requirement as the demand that curvature scalars built from polynomials of Riemann tensor components are finite. This condition corresponds to the absence of essential scalar curvature singularities. More stringent conditions that involve higher covariant derivatives or regularity of individual components are not imposed.

In a general four-dimensional spacetime, there are 14 algebraically independent scalar invariants  that can be constructed from the Riemann tensor \cite{exact:03}. We use two quantities that are expressed straightforwardly from components of the EMT, namely
\begin{align}
	\tilde{\mathrm{T}}\defeq \tensor{T}{^\mu_\mu}, \qquad \tilde{\maT} \defeq \tensor{T}{_\mu_\nu} \tensor{T}{^\mu^\nu} .
	\label{eq:scalars}
\end{align}
The Einstein equations relate them to the curvature scalars as  $\tilde{\mathrm{T}} \equiv - {\eR}/8\pi$ and $\tilde{\mathfrak{T}}\equiv \tensor{R}{^\mu^\nu}\tensor{R}{_\mu_\nu} / 64 \pi^2$, where $\tensor{R}{_\mu_\nu}$ and $\eR$ are the Ricci tensor and Ricci scalar, respectively. In spherical symmetry, if the two scalars of \eref{eq:scalars} are finite, then the rest are finite as well \cite{t:19}.

It is convenient to introduce
\begin{align}
	\tensor{\tau}{_t} \defeq e^{-2h} \tensor{T}{_t_t}, \qquad \tensor{\tau}{_t^r} \defeq e^{-h} \tensor{T}{_t^r}, \qquad \tensor{\tau}{^r} \defeq \tensor{T}{^r^r} .
	\label{eq:effectiveEMT}
\end{align}
The three Einstein equations for $\tensor{G}{_t_t}$, $\tensor{G}{_t^r}$, and $\tensor{G}{^r^r}$ can then be written as
\begin{align}
	\partial_r C &= 8 \pi r^2 \tensor{\tau}{_t} / f , \label{gtt} \\
	\partial_t C &= 8 \pi r^2 e^h \tensor{\tau}{_t^r} , \label{gtr} \\
	\partial_r h &= 4 \pi r \left( \tensor{\tau}{_t} + \tensor{\tau}{^r} \right) / f^2 . \label{grr}
\end{align}
Since $\tensor{T}{^\theta_\theta} \equiv \tensor{T}{^\varphi_\varphi}$ in spherical symmetry, the two scalars have the form
\begin{align}
	\tilde{\mathrm{T}} = \mathrm{T} + 2 \tensor{T}{^\theta_\theta}, \qquad \tilde{\maT} = \maT + 2 \big(\tensor{T}{^\theta_\theta}\big)^2,\label{twoscal0}
\end{align}
where
\begin{align}
     \mathrm{T}\defeq\left( \tensor{\tau}{^r} - \tensor{\tau}{_t} \right) / f, \qquad \mathfrak{T} \defeq \left( (\tensor{\tau}{^r})^2 + (\tensor{\tau}{_t} )^2 - 2 (\tensor{\tau}{_t^r})^2 \right) / f^2. \label{twoscal}
\end{align}
The curvature scalars are finite even if the individual terms are divergent in one of the two possible cases. The three effective EMT components $\tensor{\tau}{_a}$, $a \in \lbrace \tensor{}{_t}, \tensor{}{_t^r}, \tensor{}{^r} \rbrace$, may result in finite $\mathrm{T}$ and $\maT$, with $\tensor{T}{^\theta_\theta}$ [$\tensor{G}{^\theta_\theta}/8\pi$ with the metric functions that solve Eqs.~\eqref{gtt}--\eqref{grr}] being finite as well, or all four non-zero components $\tensor{T}{^\mu_\nu}$ may diverge, but these divergences cancel to produce the finite scalars of Eq.~\eqref{twoscal0}. We first describe the former possibility and then explain why the latter possibility is not realized.

\textit{A priori}, to ensure that $\mathrm{T}$ and $\maT$ remain finite as $r \to r_\sg$, the effective EMT components $\tensor{\tau}{_a}$ must either diverge according to the same power law $f^k$, $k<0$, converge to finite limits ($k=0$), slowly converge to zero ($0<k<1$), or converge to zero with a variety of possible exponents $k_a$, $k_a \geqslant 1$. It was shown that only two classes of dynamic solutions (with the leading terms in the functions $\tensor{\tau}{_a}$ scaling as $f^k$, $k=0,1$) satisfy the regularity conditions \cite{t:20,mt:21a}. The component $\tensor{T}{^\theta_\theta}$ is either finite due to the Einstein equations (or, possibly, due to some additional constraints on the functions $C$ and $h$), resulting in admissible solutions, or divergent, indicating the inadmissibility of the solution as $\tensor{T}{^\theta_\theta}$ cannot have the same leading divergence as $\tensor{\tau}{_a}/f$ \cite{t:20}. Moreover, a slow divergence (or a faster convergence of either $\tensor{\tau}{_t}$ or $\tensor{\tau}{^r}$) is incompatible with Eqs.~\eqref{gtr} and \eqref{grr}. Hence finite values of the two functions of Eq.~\eqref{twoscal} are a necessary condition to ensure that the curvature scalars are finite. We now briefly summarize the properties of the admissible solutions.

\subsection{Generic solution ($k=0$)} \label{generic}
In principle, the solutions with $k=0$ allow for $\tensor{\tau}{_t} \to \tensor{\tau}{^r} \to \mp \Upsilon^2$, $\tensor{\tau}{_t^r} \to \mp \Upsilon^2$ for some $\Upsilon(t)>0$, but only
\begin{align}
	& \tensor{\tau}{_t}  \approx \tensor{\tau}{^r} = - \Upsilon^2 + \mathcal{O}(\sqrt{x}), \label{eq:taut+r} \\
	& \tensor{\tau}{_t^r} = \mp \Upsilon^2 + \mathcal{O}(\sqrt{x}), \label{eq:taus}
\end{align}
where $x\defeq r-r_\sg$ denotes the coordinate distance from the horizon, describe valid solutions: taking $\tensor{\tau}{_t} \to \tensor{\tau}{^r} \to + \Upsilon^2$ results in complex-valued solutions of Eqs.~\eqref{gtt}--\eqref{grr} (see Ref.~\citenum{bmmt:19} for details).

The negative sign in the leading term of $\tensor{\tau}{_t}$ and $\tensor{\tau}{^r}$ leads to the violation of the null energy condition (NEC) \cite{he:book:73,mmv:17,ks:20} in the vicinity of the apparent horizon, i.e.\ a future-directed outward (inward) pointing radial null vector $\ell^\mu$ does not satisfy $T_{\mu\nu}\ell^\mu \ell^\nu \geqslant 0$  for the contracting (expanding) Schwarzschild radius $r_\sg$ \cite{bmmt:19}. An immediate consequence of this result is that accreting Vaidya black hole solutions in $(v,r)$ coordinates cannot describe PBHs as they satisfy the NEC \cite{bmmt:19}.

The metric functions that solve Eqs.~\eqref{gtt} and \eqref{grr} are
\begin{align}
		C &= r_\sg-4\sqrt{\pi} r_\sg^{3/2}\Upsilon\sqrt{x} + \mathcal{O}(x) , \label{eq:k0C} \\
		h &= -\frac{1}{2}\ln{\frac{x}{\xi}} + \mathcal{O}(\sqrt{x}) , \label{eq:k0h}
\end{align}
where $\xi(t)$ is determined by the choice of time variable and the higher-order terms depend on the higher-order terms in the EMT expansion \cite{bmt:19}. \eref{gtr} must then hold identically. Both sides contain terms that diverge as $1/\sqrt{x}$, and their identification results in the consistency condition
\begin{align}
	r'_\sg / \sqrt{\xi} = 4 \varepsilon_\pm \sqrt{\pi r_\sg} \, \Upsilon , \label{rpr1}
\end{align}
which describes the expansion or contraction of the Schwarzschild sphere depending on the sign of $\varepsilon_\pm = \pm 1$.

Useful information can be obtained by working with retarded and advanced null coordinates that result in regular metric functions \cite{bmmt:19}. If $r'_\sg>0$, this is achieved by using the retarded null coordinate $u$. For $r'_\sg<0$, the advanced null coordinate $v$ leads to
\begin{align}
    C_+(v,r) & = r_+(v) + \sum\limits_{i \geqslant 1}^\infty w_i(v)(r-r_+)^i,  \label{cv1} \\
    h_+(v,r) & = \sum\limits_{i \geqslant 1}^\infty \chi_i(v)(r-r_+)^i, \label{hv1}
 \end{align}
for some functions $w_i(v)$, $\chi_i(v)$, where $w_1 \leqslant 1$ due to the definition of $r_+$. This is the general form of the metric functions in $(v,r)$ coordinates that ensures finite curvature scalars at the apparent horizon if $r'_\sg<0$ \cite{t:20}. The components of the EMT in $(v,r)$ and $(t,r)$ coordinates are related by
\begin{align}
	& \theta_v \defeq e^{-2h_+} \Theta_{vv} = \tensor{\tau}{_t} , \label{thev}\\
	& \theta_{vr} \defeq e^{-h_+} \Theta_{vr} = \left( \tensor{\tau}{_t^r} - \tensor{\tau}{_t} \right) / f , \label{thevr}\\
	& \theta_r \defeq \Theta_{rr} = \left( \tensor{\tau}{^r} + \tensor{\tau}{_t} - 2 \tensor{\tau}{_t^r} \right) / f^2,   \label{ther}
\end{align}
where $\Theta_{\mu\nu}$ labels the EMT components in $(v,r)$ coordinates, which are finite.  Hence $r_\sg'<0$ corresponds to an evaporating PBH. Similarly, $(u,r)$ coordinates are regular at the anti-trapping horizon $r_\sg$ of an expanding white hole, $r_\sg'>0$ \cite{bmmt:19,mmt:rev:21}.

The limiting form of the $(tr)$ block of the EMT as $r\to r_\sg$ is
\begin{align}
	\tensor{T}{^a_b} = \begin{pmatrix}
		\Upsilon^2/f & -\varepsilon_\pm e^{-h}\Upsilon^2/f^2 \vspace{1mm}\\
		\varepsilon_\pm e^h  \Upsilon^2 & -\Upsilon^2/f
	\end{pmatrix}, \qquad
 	\tensor{T}{_{\hat{a}}_{\hat{b}}} = \frac{\Upsilon^2}{f} \begin{pmatrix}
		-1 & \varepsilon_\pm \\
		\varepsilon_\pm & -1
	\end{pmatrix},
  \label {tneg}
\end{align}
where the second expression is written in the orthonormal frame. It makes the violation of the NEC particularly transparent.

In the test-field limit \cite{bd:book:82} quantum fields propagate on a given gravitational background, but the resulting EMT is not permitted to backreact on the geometry via the Einstein equations. It is instructive to compare the  tensor of \eref{tneg} with explicit results obtained in the test-field limit. Out of the three popular choices for the vacuum state \cite{bd:book:82,fn:book:98}, only the Unruh vacuum results in an EMT with nonzero $\tensor{T}{_t_r}$ components. The state itself corresponds to the requirement that no particles impinge on the collapsing object from infinity \cite{u:76}. In the context of a static maximally extended spacetime, its counterpart is a state with unpopulated modes at past null infinity and the white hole horizon \cite{bd:book:82,fn:book:98}. Using various semi-analytical and numerical methods that are based on conformally coupled fields \cite{v:97} and minimally coupled scalar field \cite{lo:16}, the expectation values of the renormalized components $\tensor{T}{_t_t}$, $\tensor{T}{_t^r}$, and $\tensor{T}{^r^r}$ have been determined explicitly. They approach the same negative value as $r \to r_\sg$.

The experiences of observers in the vicinity of the apparent horizon depend on their trajectories. A static observer finds that the energy density $\rho \defeq \tensor{T}{_\mu_\nu} u^\mu u^\nu = - \tensor{T}{^t_t}$, pressure $p \defeq \tensor{T}{_\mu_\nu} n^\mu n^\nu = \tensor{T}{^r_r}$, and flux $\phi \defeq  \tensor{T}{_\mu_\nu} u^\mu n^\nu$ diverge at the apparent horizon, where $u^\mu$ denotes the four-velocity and $n^\mu$ the outward-pointing radial spacelike vector. The experience of a radially-infalling observer Alice moving on the trajectory $x^\mu_\mathrm{A}(\tau) = (t_\mathrm{A}, r_\mathrm{A},0,0)$ is different, and also differs from the infall into a classical eternal black hole.

First, crossing the apparent horizon of a PBH happens not only at some finite proper time $\tau_0$, but also at a finite time $t_0(\tau_0)$, $r_\sg\big(t_0(\tau_0)\big)=r_\mathrm{A}(\tau_0)$ according to the clock of a distant Bob. This is particularly easy to see for radial null geodesics, where
\begin{align}
	\frac{dt}{dr}= - \frac{e^{-h(t,r)}}{f(t,r)} \to \pm\frac{1}{|r_\sg'|}
	\label{eq:null_geodesics}
\end{align}
at $r=r_\sg$, the rhs is obtained using Eqs.~\eqref{eq:k0C}--\eqref{rpr1} \cite{t:20}, and the upper (lower) signature corresponds to an outgoing (ingoing) null ray. We use this result in our estimate of the formation time in Sec.~\ref{disc}, and in showing that the simple generalizations of surface gravity to nonstationary spacetimes fail for PBHs (Sec.~\ref{surfg}).

It is possible to find explicit relations between $(t,r)$ and $(v,r)$ coordinates in the vicinity of the apparent horizon \cite{mt:21a}, namely
\begin{align}
	x(r_++y,v) =r_++y-r_\sg\big(t(v,r_++y)\big)=  -r''_\sg y^2/(2 r_\sg'{}^2)+\cO( y^3) , \label{xyrel}
\end{align}
which relates the coordinates $x \defeq r - r_\sg(t)$ and $y(r,v) \defeq r - r_+(v)$. Another useful relation \cite{mmt:21} is
\begin{align}
	w_1 = 1 - 2 \sqrt{2\pi r_\sg^3|r_\sg''|}\frac{\Upsilon}{|r_\sg'|}. \label{w1t}
\end{align}
For an evaporating PBH ($r_\sg'<0$) the energy density, pressure, and flux in Alice's frame are finite. However, upon approaching the anti-trapping horizon of an expanding white hole ($r_\sg'>0$), Alice encounters a firewall,
\begin{align}
	\rho_{\mathrm{A}} = \tensor{T}{_\mu_\nu} u^\mu_\mathrm{A} u^\nu_\mathrm{A} = - \frac{  \dot{r}_\mathrm{A}^2}{4 \pi r_\sg X} + \mathcal{O}(1/\sqrt{X}), \label{fire}
\end{align}
where $X \defeq r_A(\tau)-r_\sg\big(t_\mathrm{A}(\tau)\big)$ \cite{t:19}, which may be another indication of the instability \cite{fn:book:98} of such objects.

Violations of the NEC are bounded by quantum energy inequalities \cite{ks:20,f:17}. Outside of the singularities the lower bounds were shown to exist for the energy density $\langle \tensor{\hat{T}}{_\mu_\nu} \rangle_\omega u^\mu u^\nu$ and its smeared averages. These are known to be state-independent for free fields. For spacetimes of small curvature, explicit bounds for a geodesic observer were derived in Ref.~\citenum{ko:15}. A finite bound is violated by the $1/f^2$ divergence of the energy density that results in the logarithmic divergence of its smeared time average provided that the radial velocity does not go to zero \cite{t:20,dmt:21}. However, this is not true for geodesics, and thus the inequality of Ref.~\citenum{ko:15} is inapplicable.

Observers that have crossed the apparent horizon but change their mind before traversing the quantum ergosphere can exit the black hole before it evaporates. Their experiences at the apparent horizon are more involved: for a geodesic observer that attempts to cross the apparent horizon from the inside, the energy density also diverges. However, it does so according to a weaker $1/f$ law, and thus the integrated energy density remains finite.

\subsection{Extreme solution ($k=1$)}
The static solution with $k=0$ is impossible, as in this case $\mathfrak{T}$ would diverge at the apparent horizon. Consequently, EMT components that allow for static solutions must behave differently. Many models of static nonsingular black holes assume finite values of energy density and pressure at the horizon \cite{f:14,h:06}. With respect to the  scaling behavior $f^k$ of the effective EMT components of \eref{eq:effectiveEMT} this is the $k=1$ solution, with
\begin{align}
	\tensor{\tau}{_t} \to E(t)f, \qquad \tensor{\tau}{_t^r} \to \Phi(t)f, \qquad \tensor{\tau}{^r} \to P(t)f ,
	\label{eq:k1taus}
\end{align}
where $\rho=E$ and $p=P$ at the apparent horizon. Any two functions can be expressed algebraically in terms of the third, and $8 \pi r_\sg ^2E \leqslant 1$ to ensure that $C(t,r)-r_\sg>0$ for $r>r_\sg$.

Only the extreme value of $E=(8\pi r_\sg^2)^{-1}$ corresponds to non-extreme dynamic black holes (i.e.\ those with $r_\sg \neq \mathrm{const}$ whose trapped regions have nonzero volume) \cite{mt:21a}. As a result
\begin{align}
	C(t,r)=r- c_{32}(t)x^{3/2}+\cO(x^2) \label{fk1}
\end{align}
for some coefficient $c_{32}(t)>0$, setting via \eref{eq:k1taus} the scaling of other leading terms in the EMT. Consistency of Eqs.~\eqref{gtr} and \eqref{grr} implies $P=-E=-1/(8\pi r^2_\sg)$ and $\Phi=0$. From the next-order expansion we obtain
\begin{align}
	h = - \frac{3}{2}\ln \frac{x}{\xi} + \mathcal{O}(\sqrt{x}), \label{hk1}
\end{align}
as well as the consistency relation
\begin{align}
	r'_\sg = \varepsilon_\pm c_{32}\xi^{3/2}/r_\sg , \label{rate1}
\end{align}
where the case $\varepsilon_\pm \equiv -1$ ($\varepsilon_\pm \equiv +1$) describes an evaporating PBH (an expanding white hole). The NEC violation is more subtle than in the $k=0$ case. At $r=r_\sg(t)$ itself the NEC is marginally satisfied, as
\begin{align}
 	\tensor{T}{_{\hat{a}}_{\hat{b}}} = \frac{1}{8\pi r_g^2} \begin{pmatrix}
		1 & 0 \\
		0 & -1
	\end{pmatrix} . \label{tneg1}
\end{align}
However, the NEC is violated for some $x>0$, as for both incoming and outgoing directions $l_\mathrm{in}$, $l_\mathrm{out}$, we have
\begin{align}
	\tensor{T}{_{\hat{a}}_{\hat{b}}} l^{\hat a} l^{\hat b} = - \frac{3 c_{32} \sqrt{x}}{8 \pi r_\sg^2} + \mathcal{O}(x) .
\end{align}
Solutions with a time-independent apparent horizon or general static solutions do not require $w_1=1$ to satisfy Eqs.~\eqref{thev}--\eqref{ther}. Since $r_+(v)=r_\sg(t)=\mathrm{const}$, it is possible to have non-extreme solutions. Then \eref{gtr} implies $\Phi=0$ and the identity $E=-P$ follows from \eref{ther}, leading to a regular function $h(t,r)$. However, in this case \eref{intf} indicates that the apparent horizon cannot be reached in finite time $t$.

The curvature scalars of the solutions we have considered so far were rendered finite thanks to the constraint that Eq.~\eqref{gtr} imposes on the dynamic behavior via Eqs.~\eqref{rpr1} and \eqref{rate1}. For example, the Ricci scalar of the solution with $C(t,r)$ given by Eq.~\eqref{fk1} includes the potentially divergent term
\begin{align}
	R_\mathrm{div}=\frac{3}{4\sqrt{x}}\left(\frac{r_\sg r_\sg'^2}{c_{32}\xi^3}- \frac{c_{32}}{r_\sg}\right)
\end{align}
that vanishes upon substitution of Eq.~\eqref{rate1}. However, a metric with $C(t,r)=r+c_2x^2+\ldots$ and the same $h(t,r)$ of Eq.~\eqref{hk1}, while being a solution of the Einstein equations, has a finite value of the Ricci scalar only in the static case $r_\sg'=0$, as in this case
\begin{align}
	R_\mathrm{div}=-\frac{3 r_\sg r_\sg'^2}{c_2\xi^3x}.
\end{align}

\section{Implications}
The solutions described in \sref{metrics} have many remarkable properties,  {and their near-horizon geometry differs considerably from that of Schwarzschild black hole spacetimes}. First, we mention an immediate mathematical consequence of the firewall that we have described. The divergence of $\rho_\mathrm{A}$ [see \eref{fire}] points to the presence of a matter singularity. As a result, we observe that if an apparent/anti-trapping horizon forms, it is a surface with an intermediate singularity. The appearance of a negative energy firewall is the counterpart to arbitrarily large tidal forces that could tear apart an observer falling into such a singularity. In these cases the fate of an observer depends on the integrated tidal stress \cite{ek:74,fn:book:98}.

Consider observers attempting to enter an expanding white hole or exit an evaporating PBH \cite{mmt:rev:21}. In the former case, the radial velocity $\dot R<0$ on a geodesic trajectory reverses the sign at some finite $R-r_\sg$ due to the divergent radial acceleration, and in fact the energy density remains finite. On the other hand, a massive test particle exits a PBH with $\dot R=0$ and the energy density diverges, but its integrated smeared version remains finite \cite{dmt:21}.

Since all curvature scalars remain finite, it is instructive to check the Ricci spinors
\begin{align}
	\Phi_{00} = \tfrac{1}{2} R_{11}, \quad \Phi_{22} = \tfrac{1}{2} R_{22}, \quad \Phi_{11} = \tfrac{1}{4}(R_{12}+R_{34}) .
\end{align}
Using the natural Newman--Penrose tetrad that is built from the two null vectors of \eref{null-v} and a pair of complex-conjugate vectors $m^\mu$ and $\bar m^\mu \defeq m^{\mu*}$,
\begin{align}
	m=\frac{1}{\sqrt{2}r}\pad_\theta+\frac{i}{\sqrt{2}r\sin\theta}\pad_\phi, \qquad m\cdot \bar m=1,
\end{align}
we find that the values of all nonzero spinors are finite at the apparent horizon. However, given the freedom of choice of these vectors $l^\mu_{\mathrm{out}}\to Al^\mu_{\mathrm{out}}$, $l^\mu_{\mathrm{in}}\to l^\mu_{\mathrm{in}}/A$, the values of the spinors $\Phi_{00}$ and $\Phi_{22}$ depend on this choice. By choosing $A=f(v,r)$ (this form of the tangent vectors may appear more natural in $(t,r)$ coordinates), we find
\begin{align}
	\Phi_{00}\propto f, \qquad \Phi_{22}\propto f^{-1},
\end{align}
again demonstrating that the apparent/anti-trapping horizon is a surface of intermediate singularity \cite{dmt:21}.

\subsection{Black hole formation} \label{sec:PBHform}
Consider now possibilities for horizon formation. The properties of the self-consistent solutions of the Einstein equations that we have described above lead to the identification of a unique scenario for black hole formation \cite{t:20,mt:21a}. We consider only evaporating PBHs ($r'_\sg<0$) due to the implications of the results presented in Sec.~\ref{generic}. Working in $(v,r)$ coordinates, we assume that the first marginally trapped surface appears at some $v_\mathrm{S}$ at $r_+(v_\rS)$ that corresponds to a finite value  of $t_\rS$. For $v \leqslant \vS$, the MS mass in its vicinity can be described by modifying \eref{cv1} as
\begin{align}
	C(v,r) = \Delta(v) + r_*(v) + \sum\limits_{i \geqslant 1}^\infty w_i(v) (r-r_*)^i ,
\end{align}
where  $r_*(v)$ corresponds to the maximum of $\Delta_v(r)\defeq C(v,r) - r$, and the deficit function $\Delta(v) \defeq C(v,r_*)-r_*(v) \leqslant 0$. At the advanced time $\vS$, the location of the maximum corresponds to the first marginally trapped surface, $r_*(\vS) = r_+(\vS)$, and $\sigma(\vS)=0$. For $v \geqslant \vS$, the MS mass is described by \eref{cv1}. For $v \leqslant \vS$, the (local) maximum of $\Delta_v(r)$ is determined by $\Delta_v(r)/dr=0$, hence $w_1(v) - 1 \equiv 0$.

There are no \textit{a priori} restrictions on the evolution of $r_*$ before the PBH is formed. However, since an expanding Schwarzschild sphere  describes white hole solutions $r'_+(\vS) < 0$. For $v>\vS$, the maximum of $C(v,r)$ does not coincide with $r_+(v)$ since the trapped region is of finite size. As a result, $w_1(v)<1$ for $v>\vS$.

This means that at its formation a PBH is described by a $k=1$ solution, which can be seen from \eref{w1t}, as $w_1=1$ implies $\Upsilon=0$. It then immediately switches to the $k=0$ solution, with matching decrease in $w_1(v)$ and increase in $\Upsilon \big(t(v,r_+)\big)$, and $w_1<1$ at all subsequent stages. The transition from $f^1$ to $f^0$ behavior is continuous as $\Upsilon(t_\mathrm{S}) \equiv 0$ for the $k=1$ solution and it increases thereafter \cite{mt:21a}.  A detailed discussion can be found in Ref.~\citenum{mmt:rev:21}.

\subsection{Surface gravity} \label{surfg}
The surface gravity $\kappa$ plays an important role in GR and semiclassical gravity \cite{he:book:73,fn:book:98,f:book:15}. However, it is unambiguously defined only in stationary spacetimes (for a Schwarzschild black hole $\kappa=(2r_\sg)^{-1}$), where it is proportional to the Hawking temperature.

Generalizations of surface gravity to dynamical spacetimes are related to two {(equivalent in stationary spacetimes)} definitions of $\kappa$ that are based on the peeling off properties and the inaffinity of null geodesics at the horizon, respectively \cite{f:book:15,clv:13}. For sufficiently slowly evolving horizons with properties sufficiently close to their classical counterparts, these different generalizations of surface gravity are practically indistinguishable \cite{clv:13}. This agreement is important. Emission of Hawking-like radiation does not require the formation of an event or even an apparent horizon \cite{pp:09,h:87,blsv:06,vsk:07}. The role of the Hawking temperature is captured in various derivations either by the peeling \cite{blsv:11} or the Kodama \cite{kpv:21} surface gravity.

For PBHs, however, these quantities are very different. Here we consider only one version of the peeling surface gravity \cite{mmt:21}. In Schwarzschild coordinates  $\kappa_\mathrm{peel}$ is defined from the near-horizon behavior of null geodesics \cite{f:book:15,clv:13} as the linear coefficient in the Taylor expansion of
\begin{align}
	\frac{dr}{dt}=\pm e^{h}f(t,r)
\end{align}
in powers of $x$ as $x \to 0$ (i.e.\ $r \to r_\sg$). However, such an expansion is impossible for the metric functions of Eqs.~\eqref{cv1}--\eqref{hv1}. Alternatively, to be compatible with \eref{eq:null_geodesics}, $\kappa_\mathrm{peel}$ should be infinite, as we now demonstrate.

For differentiable $C$ and $h$, the result is
\begin{align}
	\kappa_\mathrm{peel} = \frac{e^{h(t,r_\sg)} \left( 1 - C'(t,r_\sg) \right)}{2 r_\sg} .
\end{align}
This quantity is undefined for $k=0$ solutions as for a radial geodesic [\eref{eq:null_geodesics}] it implies
\begin{align}
	\frac{dr}{dt} = \pm|r_\sg'| + \mathcal{O}(\sqrt{x}),
\end{align}
and $\kappa_\mathrm{peel}$ diverges (for both $k=0$ and $k=1$ solutions) \cite{mt:21a}.

The Kodama vector field has many useful properties of the Killing field to which, modulo possible rescaling, it reduces in the static case \cite{f:book:15,clv:13,av:10}. Similar to the Killing vector, it is most conveniently expressed in $(v,r)$ coordinates,
\begin{align}
	K^\mu=(e^{-h_+},0,0,0).
\end{align}
It is covariantly conserved, and generates the conserved current
\begin{align}
	\nabla_\mu K^\mu &= 0 , \\
	\nabla_\mu J^\mu &= 0 , \qquad J^\mu \defeq \tensor{G}{^\mu^\nu} K_\nu ,
\end{align}
thereby giving a natural geometric meaning to the Schwarzschild coordinate time $t$. The MS mass is its Noether charge.

Since $K_{(\mu;\nu)}\neq 0$, the generalized Hayward--Kodama surface gravity is defined via \cite{h:98}
\begin{align}
	\frac{1}{2} K^\mu(\nabla_\mu K_\nu-\nabla_\nu K_\mu) \defeq \kappa_\mathrm{K} K_\nu,
\end{align}
evaluated on the apparent horizon. Hence
\begin{align}
	\kappa_\mathrm{K} = \frac{1}{2} \left.\left(\frac{C_+(v,r)}{r^2} - \frac{\partial_r C_+(v,r)}{r}\right) \right|_{r=r_+} \hspace*{-3mm} = \frac{(1-w_1)}{2r_+},
\end{align}
where Eq.~\eqref{cv1} was used to obtain the final result. Thus at the formation of a black hole (i.e.\ of the first trapped surface) this version of surface gravity is zero. At the subsequent evolution stages that correspond to a $k=0$ solution, $\kappa_\mathrm{K}$ is nonzero. However, it approaches the static value $\kappa=1/(4M)$ only if the metric is close to the pure Vaidya metric with $w_1\equiv 0$. This in turn contradicts the semiclassical results \cite{mmt:21}.

\section{Discussion} \label{disc}
The solutions we have described follow from commonly obtained or assumed, but not always sufficiently articulated requirements: the formation of a minimally regular horizon in finite time of a distant observer. The two classes of spherically symmetric solutions that satisfy these requirements must violate the NEC in the vicinity of the outer apparent horizon, and the matter content is dominated by a null fluid. Unless we accept that semiclassical physics breaks down at the horizon scale, accretion is no longer possible after the horizon has formed: the solutions describe either evaporating black holes or expanding white holes.

It is still not clear how the collapsing matter actually behaves, and how the continuous transition from $f^1$ to $f^0$ behavior immediately after PBH formation (\sref{sec:PBHform}) can be realized in nature. A mechanism that converts the original matter into the exotic matter present in the vicinity of the forming apparent horizon is required to violate the NEC. However, the emission of collapse-induced radiation \cite{h:87,blsv:06,vsk:07} is a nonviolent process that approaches at latter times the standard Hawking radiation and Page's evaporation law \cite{fn:book:98,p:76} $r_\sg' = - \alpha/r_\sg^2$, $\alpha \sim 10^{-3}-10^{-4}$.

In addition, there is no obvious way to reconcile the two standard definitions of surface gravity that underpin different derivations of Hawking radiation to dynamical spacetimes (\sref{surfg}): the peeling surface gravity $\kappa_\mathrm{peel}$ diverges, whereas the Kodama surface gravity $\kappa_\mathrm{K}$ is zero at the instant of PBH formation and in contradiction with established semiclassical results thereafter. This has interesting repercussions for the formulation of the information loss paradox \cite{mmt:21} and indicates that the Hawking radiation that is obtained on the Schwarzschild background is not an asymptotic form of the radiation that is emitted by a PBH.

Our results indicate that the observed ABHs may be horizonless UCOs rather than genuine PBHs with a horizon --- not due to some exotic supporting matter or dramatic variation in the laws of gravity, but simply because the conditions for the formation of a horizon have not been met at the present moment of $t$. Even if no new physics is required to produce the mandatory NEC violation, the process may be too slow to transform the UCOs that we observe into PBHs. Eq.~\eqref{eq:null_geodesics} of Sec.~\ref{generic} sets the timescale of the last stages of infall according to Bob. Assuming that it is applicable through the radial interval of the order of $r_\sg$, we find $t_\mathrm{in}\sim r_\sg/r'_\sg$. For an evaporating macroscopic PBH, this is of the same order of magnitude as the Hawking process decay time $t_\mathrm{evp}\sim 10^3 r_\sg^3$. Such behavior was found in thin shell collapse models, where the exterior geometry is modeled by a pure outgoing Vaidya metric \cite{bmt:19}. For a solar mass black hole, this time is about $10^{67}$yr, indicating that it is simply too early for the horizon to form. It is also conceivable that the conditions are not met before evaporation is complete or before effects of quantum gravity become dominant \cite{fn:book:98}. Horizon avoidance may therefore occur due to the absence of new physics, and not because of it.

A more detailed understanding of the near-horizon geometry as well as the behavior of matter during gravitational collapse and PBH formation is needed to resolve all of the remaining questions. Coming of age of multimessenger astronomy and controversies surrounding the direct observation of ABHs make it a particularly timely project.

\section*{Acknowledgments}
SM is supported by an International Macquarie University Research Excellence Scholarship and a Sydney Quantum Academy Scholarship. The work of DRT is supported by the ARC Discovery project grant DP210101279.


\begin{thebibliography}{99}
	\bibitem{map:19} L.\ Barack, V.\ Cardoso, S.\ Nissanke, and T.\ P.\ Sotiriou (eds.), {\href{https://doi.org/10.1088/1361-6382/ab0587}{Class.\ Quantum Gravity \textbf{36}, 143001 (2019)}}.
	\bibitem{cp:19} V.\ Cardoso and P.\ Pani, {\href{https://doi.org/10.1007/s41114-019-0020-4}{Living Rev.\ Relativ.\ \textbf{22}, 4 (2019)}}.
	\bibitem{c:19} E.\ Curiel, {\href{https://doi.org/10.1038/s41550-018-0602-1}{Nat.\ Astron.\ \textbf{3}, 27 (2019)}}.
	\bibitem{f:14} V.\ P.\ Frolov, {\href{https://doi.org/10.1007/JHEP05(2014)049}{J.\ High Energy Phys.\ \textbf{05}, 049 (2014)}}.
	\bibitem{v:14} M.\ Visser, {\href{https://doi.org/10.1103/PhysRevD.90.127502}{Phys.\ Rev.\ D \textbf{90}, 127502 (2014)}}.
	\bibitem{cp:17} V.\ Cardoso and P.\ Pani, {\href{https://doi.org/10.1038/s41550-017-0225-y}{Nat.\ Astron.\ \textbf{1}, 586 (2017)}}.
	\bibitem{he:book:73} S.\ W.\ Hawking and G.\ F.\ R.\ Ellis, {\href{https://doi.org/10.1017/CBO9780511524646}{\textit{The Large Scale Structure of Space-Time} (Cambridge University Press, Cambridge, England, 1973)}}.
	\bibitem{fn:book:98} V.\ P.\ Frolov and I.\ D.\ Novikov, {\href{https://doi.org/10.1007/978-94-011-5139-9}{\textit{Black Hole Physics: Basic Concepts and New Developments} (Kluwer, Dordrecht, 1998)}}.
	\bibitem{f:book:15} V.\ Faraoni, {\href{https://doi.org/10.1007/978-3-319-19240-6}{\textit{Cosmological and Black Hole Apparent Horizons} (Springer, Heidelberg, 2015)}}.
	\bibitem{mmt:rev:21} R.\ B.\ Mann, S.\ Murk, and D.\ R.\ Terno, {\href{https://doi.org/10.1142/S0218271822300154}{Int.\ J.\ Mod.\ Phys.\ D \textbf{31}, 2230015 (2022)}}.
	\bibitem{dmt:21} P.\ K.\ Dahal, S.\ Murk, and D.\ R.\ Terno, {\href{https://doi.org/10.1116/5.0073598}{AVS Quantum Sci.\ \textbf{4}, 015606 (2022)}}.	
	\bibitem{h:75} S.\ W.\ Hawking, {\href{https://doi.org/10.1007/BF02345020}{Commun.\ Math.\ Phys.\ \textbf{43}, 199 (1975)}}.
	\bibitem{y:83} J.\ W.\ York, Jr., {\href{https://doi.org/10.1103/PhysRevD.28.2929}{Phys.\ Rev.\ D \textbf{28}, 2929 (1983)}}.
	\bibitem{cdlv:20} R.\ Carballo-Rubio, F.\ Di Filippo, S.\ Liberati, and M.\ Visser,  {\href{https://doi.org/10.1103/PhysRevD.101.084047}{Phys.\ Rev.\ D \textbf{101}, 084047 (2020)}}.
	\bibitem{bhl:18} P.\ Bin\'{e}truy, A.\ Helou and F.\ Lamy, {\href{https://doi.org/10.1103/PhysRevD.98.064058}{Phys.\ Rev.\ D \textbf{98}, 064058 (2018)}}.
	\bibitem{sg:15} J.\ M.\ M.\ Senovilla and D.\ Garfinkle, {\href{https://doi.org/10.1088/0264-9381/32/12/124008}{Class.\ Quantum Gravity \textbf{32}, 124008 (2015)}}.
	\bibitem{c-b:09} Y.\ Choquet-Bruhat, {\href{https://doi.org/10.1093/acprof:oso/9780199230723.001.0001}{\textit{General Relativity and the Einstein Equations} (Oxford University Press, Oxford, England, 2009)}}.
	\bibitem{bd:book:82} N.\ D.\ Birrel and P.\ C.\ W.\ Davies, {\href{https://doi.org/10.1017/CBO9780511622632}{\textit{Quantum Fields in Curved Space} (Cambridge University Press, Cambridge, England, 1984)}}.
	\bibitem{w:01} R.\ M.\ Wald, {\href{https://doi.org/10.12942/lrr-2001-6}{Living Rev.\ Relativ.\ \textbf{4}, 6 (2001)}}.
	\bibitem{rb:83} T.\ A.\ Roman and P.\ G.\ Bergman, {\href{https://doi.org/10.1103/PhysRevD.28.1265}{Phys.\ Rev.\ D \textbf{28}, 1265 (1983)}}.
	\bibitem{h:06} S.\ A.\ Hayward, {\href{https://doi.org/10.1103/PhysRevLett.96.031103}{Phys.\ Rev.\ Lett.\ \textbf{96}, 031103 (2006)}}.
	\bibitem{av:10} G.\ Abreu and M.\ Visser, {\href{https://doi.org/10.1103/PhysRevD.82.044027}{Phys.\ Rev.\ D \textbf{82}, 044027 (2010)}}.
	\bibitem{ms:64} C.\ W.\ Misner and D.\ H.\ Sharp, {\href{https://doi.org/10.1103/PhysRev.136.B571}{Phys.\ Rev.\ \textbf{136}, B571 (1964)}}.
	\bibitem{fefhm:17} V.\ Faraoni, G.\ F.\ R.\ Ellis, J.\ T.\ Firouzjaee, A.\ Helou, and I.\ Musco, {\href{https://doi.org/10.1103/PhysRevD.95.024008}{Phys.\ Rev.\ D \textbf{95}, 024008 (2017)}}.
	\bibitem{p:68} R.\ Penrose, \textit{Structure of space-time}, in C.\ DeWitt, J.\  A.\ Wheeler (eds.), \textit{Batelle Rencontres: 1967 Lectures in Mathematics and Physics}, pp.\ 121\textendash{}235 (W.\ A.\ Benjamin, San Francisco, 1968).
	\bibitem{tce:80} F.\ J.\ Tipler, C.\ J.\ S.\ Clarke, and G.\ R.\ F.\ Ellis, \textit{Singularities and horizons: a review article}, in A.\ Held (ed.), \textit{General relativity and gravitation: one hundred years after the birth of Albert Einstein}, vol.\ 2, p.\ 97 (Plenum, New York, 1980).
	\bibitem{exact:03} H.\ Stephani, D.\ Kramer, M.\ MacCallum, C.\ Hoenselaers, and E.\ Herlt, {\href{https://doi.org/10.1017/CBO9780511535185}{\textit{Exact Solutions of Einstein's Field Equations}, 2nd ed.\ (Cambridge University Press, Cambridge, England, 2003)}}.
	\bibitem{t:19} D.\ R.\ Terno, {\href{https://doi.org/10.1103/PhysRevD.100.124025}{Phys.\ Rev.\ D \textbf{100}, 124025 (2019)}}.
	\bibitem{t:20} D.\ R.\  Terno, {\href{https://doi.org/10.1103/PhysRevD.101.124053}{Phys.\ Rev.\ D \textbf{101}, 124053 (2020)}}.
	\bibitem{mt:21a} S.\ Murk and D.\ R.\ Terno, {\href{https://doi.org/10.1103/PhysRevD.103.064082}{Phys.\ Rev.\ D \textbf{103}, 064082 (2021)}}.
	\bibitem{bmmt:19} V.\ Baccetti, R.\ B.\ Mann, S.\ Murk, and D.\ R.\ Terno, {\href{https://doi.org/10.1103/PhysRevD.99.124014}{Phys.\ Rev.\ D \textbf{99}, 124014 (2019)}}.
	\bibitem{mmv:17} P.\ Mart\'{\i}n-Moruno and M.\ Visser, \textit{Classical and Semi-classical Energy Conditions}, in {\href{https://doi.org/10.1007/978-3-319-55182-1}{\textit{Wormholes, Warp Drives and Energy Conditions,} edited by F.\ N.\ S.\ Lobo (Springer, New York, 2017)}}, p.\ 193.
	\bibitem{ks:20} E.-A.\ Kontou and K.\ Sanders, {\href{https://doi.org/10.1088/1361-6382/ab8fcf}{Class.\ Quantum Gravity \textbf{37}, 193001 (2020)}}.
	\bibitem{bmt:19} V.\ Baccetti, S.\ Murk, and D.\ R.\ Terno, {\href{https://doi.org/10.1103/PhysRevD.100.064054}{Phys.\ Rev.\ D \textbf{100}, 064054 (2019)}}.
	\bibitem{u:76} W.\ G.\ Unruh, {\href{https://doi.org/10.1103/PhysRevD.14.870}{Phys.\ Rev.\ D \textbf{14}, 870 (1976)}}.
	\bibitem{v:97} M.\ Visser, {\href{https://doi.org/10.1103/PhysRevD.56.936}{Phys.\ Rev.\ D \textbf{56}, 936 (1997)}}.
	\bibitem{lo:16} A.\ Levi and A.\ Ori, {\href{https://doi.org/10.1103/PhysRevLett.117.231101}{Phys.\ Rev.\ Lett.\ \textbf{117}, 231101 (2016)}}.
	\bibitem{mmt:21} R.\ B.\ Mann, S.\ Murk, and D.\ R.\ Terno, {\href{https://doi.org/10.1103/PhysRevD.105.124032}{Phys.\ Rev.\ D \textbf{105}, 124032 (2022)}}.
	\bibitem{f:17} C.\ J.\ Fewster, \textit{Quantum Energy Inequalities}, in {\href{https://doi.org/10.1007/978-3-319-55182-1}{\textit{Wormholes, Warp Drives and Energy Conditions}, edited by F.\ N.\ S.\ Lobo (Springer, New York, 2017)}}, p.\ 215.
	\bibitem{ko:15} E.-A.\ Kontou and K.\ D.\ Olum, {\href{https://doi.org/10.1103/PhysRevD.91.104005}{Phys.\ Rev.\ D \textbf{91}, 104005 (2015)}}.
	\bibitem{ek:74} G.\ F.\ R.\ Ellis and A.\ R.\ King, {\href{https://doi.org/10.1007/BF01651508}{Commun.\ Math.\ Phys.\ \textbf{38}, 119 (1974)}}.
	\bibitem{clv:13} B.\ Cropp, S.\ Liberati, and M.\ Visser, {\href{https://doi.org/10.1088/0264-9381/30/12/125001}{Class.\ Quantum Gravity \textbf{30}, 125001 (2013)}}.
	\bibitem{h:87} P. H\'{a}j\'{\i}\v{c}ek, {\href{https://doi.org/10.1103/PhysRevD.36.1065}{Phys.\ Rev.\ D \textbf{36}, 1065 (1987)}}.
	\bibitem{blsv:06} C.\ Barcel\'{o}, S.\ Liberati, S.\ Sonego, and M.\ Visser, {\href{https://doi.org/10.1088/0264-9381/23/17/014}{Class.\ Quantum Gravity \textbf{23}, 5341 (2006)}}.
	\bibitem{vsk:07} T.\ Vachaspati, D.\ Stojkovic, and L.\ M.\ Krauss, {\href{https://doi.org/10.1103/PhysRevD.76.024005}{Phys.\ Rev.\ D \textbf{76}, 024005 (2007)}}.
	\bibitem{pp:09} A.\ Paranjape and T.\ Padmanabhan, {\href{https://doi.org/10.1103/PhysRevD.80.044011}{Phys.\ Rev.\ D \textbf{80}, 044011 (2009)}}.
	\bibitem{blsv:11} C.\ Barcel\'{o}, S.\ Liberati, S.\ Sonego, and M.\ Visser, {\href{https://doi.org/10.1103/PhysRevD.83.041501}{Phys.\ Rev.\ D \textbf{83}, 041501(R) (2011)}}.
	\bibitem{kpv:21} F.\ Kurpicz, N.\ Pinamonti, and R.\ Verch, {\href{https://doi.org/10.1007/s11005-021-01445-7}{Lett.\ Math.\ Phys.\ \textbf{111}, 110 (2021)}}.	
	\bibitem{h:98} S.\ A.\ Hayward, {\href{https://doi.org/10.1088/0264-9381/15/10/017}{Class.\ Quantum Gravity \textbf{15}, 3147 (1998)}}.
	\bibitem{p:76} D.\ N.\ Page, {\href{https://doi.org/10.1103/PhysRevD.13.198}{Phys.\ Rev.\ D \textbf{13}, 198 (1976)}}.
\end{thebibliography}
\end{document}